\documentclass[12pt]{article}
\usepackage{graphicx}
\empty
\begin{document}
\pagestyle{headings}
\title{$(g-2)_{\mu}$ anomaly within the left-right symmetric
model}
\author{O. M. Boyarkin\thanks{boyarkin@front.ru}, G. G. Boyarkina and V. V. Makhnach\\
\it{Belarus State Pedagogical University}\\
\it{Soviet Street 18, Minsk, 220050, Belarus}}
\date{}
\maketitle {\small{The muon anomalous magnetic moment is discussed
within the model based on the $SU(2)_L\times SU(2)_R\times
U(1)_{B-L}$-gauge group. The contributions of the neutrinos
having the dipole magnetic moment (DMM) are calculated. The cases
of the Majorana and Dirac neutrino are considered. It is shown
that to account for the $(g-2)_{\mu}$ anomaly DMMs connected
with heavy neutrinos should be on the order of
$\mbox{few}\times10^{-8}\mu_B$, that is at variance with the theoretically
predicted values.}}

\vspace*{5mm} {\small{PACS number(s):12.15.Ff, 13.40.Em}}

\section{Introduction}

The muon anomalous magnetic moment (AMM) has been measured in a series
of three experiments at CERN (1968-1977) and, most recently, in
E-0821 experiment at the Brookhaven National Laboratory (BNL) on
Alternating Gradient Synchrotron (in 1997-2001). The results of
the CERN-III experiment were combined to yield a 7.3 ppm
measurement in agreement with a theory. The measurements at BNL,
using nearly equal samples of positive and negative muons, have
been carried out with an impressive accuracy of 0.72 ppm and
yielded the present world average \cite{Bnt}
$$a_{\mu}^{exp}=11\ 659\ 208.0(6.3)\times10^{-10}\eqno(1)$$
with an accuracy of 0.54 ppm, which represents a 14-fold
improvement compared to the previous measurements at CERN.
However, contrary to expectations, the BNL data do not agree with
the standard model (SM) prediction. For example, the result of
Ref. \cite{JPM} based on $e^-e^+$ data alone is as follows
$$a_{\mu}^{SM}=11\ 659\ 178.5(6.1)\times10^{-10}.\eqno(2)$$
Introducing the quantity
$$\delta a_{\mu}\equiv a_{\mu}^{exp}-a_{\mu}^{SM},$$
we obtain
$$\delta a_{\mu}=(29.5\pm8.8)\times10^{-10}.\eqno(3)$$
So, the SM evaluation displays 3.4 standard deviations below the
experimental result. Analogous calculations obtained in Refs.
\cite{FJ},\cite{Dav},\cite{Hag} bring to slightly different
results but all yield deviations of more than 3 $\sigma$. In what
follows we use the value $a_{\mu}^{SM}$ given by Eq. (2). Then at
90\% CL, $\delta a_{\mu}$ must lie in the range
$$18.2\times10^{-10}\leq\delta a_{\mu}\leq40.7\times10^{-10}.\eqno(4)$$

Since the BNL data have been perfectly collected and investigated
for many years, it is highly improbable that this discrepancy
could be due to a mere statistical fluctuation, as
several earlier deviations from SM turned out to be. An extremely small
variation of the muon AMM central value  present
in all the BNL results is also a serious argument in favor
of the E-0821 experiment reliability. So, the
$(g-2)_{\mu}$ anomaly may be examined as the New Physics (NP)
signal even at a weak scale. As this takes place, it is
expected that NP contributions should be proportional to
$m_{\mu}^2/M^2_{NP},$ where $M_{NP}$ is the mass of a particle
additional to the sector of the SM particles.

In this work we continue investigation of the $(g-2)_{\mu}$
anomaly within the model based on the $SU(2)_L\times SU(2)_R\times
U(1)_{B-L}$ gauge group (left-right model) \cite{Moh}. The
left-right model (LRM) contains the whole collection of particles,
which are additional with respect to the SM. In case when the
Higgs sector holds the bidoublet, the left- and right-handed
triplets
$$\Phi\left({1\over2},{1\over2},0\right),\qquad
\Delta_L(1,0,2),\qquad\Delta_R(0,1,2),\eqno(5)$$ where in brackets
the quantum numbers of the $SU(2)_L$, $SU(2)_R$ and $U(1)_{B-L}$
groups are given, after a spontaneous symmetry breaking we have 14
physical Higgs bosons. In Ref. \cite{GO1} it was shown that at the
definite parameters values of the LRM the contributions from the
physical Higgs bosons can explain the BNL results. The influence
of the $W_2$ and $Z_2$ gauge bosons has been investigated earlier
in Ref. \cite{JP}, where it was shown that the $Z_2$ contribution
is negative, while in order to explain the E-0821 result the $W_2$
gauge boson mass must lie around 100 Gev that is at variance with
the experimental data. The problem is that, besides the Higgs and
gauge bosons, the LRM contains another three additional particles:
three heavy neutrinos $N_e, N_{\mu},N_{\tau}$ --- representing a
mixture of the mass eigenstates $N_1, N_2$ and $N_3$. As it was
demonstrated in Ref. \cite{OGB}, the one of the $N_{1,2,3}$
neutrinos could have the mass lying on the electroweak scale.

The goal of this work is to study the influence of the LRM
neutrino sector on the muon AMM value. We assume that neutrinos
possess a dipole magnetic moment (DMM) and consider the cases when
the neutrinos have both the Majorana and Dirac nature. In the next
section we estimate the contributions of the diagrams with the
neutrino exchange into the $(g-2)_{\mu}$ anomaly. Comparing the
results obtained, we find the constraints on the neutrino DMM. And
in Sec. III our conclusions are summarized.

\section{Neutrino corrections to the muon AMM}

The conventional choice of the LRM Higgs sector is given by Eq.
(5). Such a choice ensures the Majorana nature of the neutrinos.
Let us start our investigation with this very case. Then, if we
are dealing with diagonal elements of the multipole moments, the
only nonvanishing multipole moment is an anapole moment. However
this is not true for nondiagonal elements. For example,
nondiagonal elements of the dipole magnetic moment could be
nonzero. Then the neutrino DMM induced by the radiative
corrections leads to the appearance of the following terms in the
effective interaction Lagrangian
$$H_{add}=\sum_{i\neq j}\left[\mu_{\nu_i\nu_j}\overline{\nu}_i(p)\sigma_{\lambda
\tau}q_{\tau}\nu_j(p^{\prime})+\mu_{N_iN_j}\overline{N}_i(p)
\sigma_{\lambda\tau}q_{\tau}N_j(p^{\prime}) +\right.$$
$$\left.+\mu_{\nu_iN_j}\overline{\nu}_i(p)\sigma_{\lambda
\tau}q_{\tau}N_j(p^{\prime})+\mu_{N_i\nu_j}\overline{N}_i(p)\sigma_{\lambda
\tau}q_{\tau}\nu_j(p^{\prime})\right]A_{\lambda}(q), \eqno(6)$$
where
$$\sigma_{\lambda\tau}={i\over2}\left(\gamma_{\tau}\gamma_{\lambda}-
\gamma_{\lambda}\gamma_{\tau}\right),\qquad
q_{\tau}=(p^{\prime}-p)_{\tau}.$$ For the sake of simplicity, we
suppose that mixing takes place between the muon and tau
neutrino only, namely:
$$\left(\matrix{\nu_{\mu L}\cr \nu_{\tau L}\cr N_{\mu R}\cr N_{\tau R}}\right)
={\cal{M}}\left(\matrix{\nu_2\cr \nu_3\cr N_2\cr
N_3}\right),\eqno(7)$$ where
$${\cal{M}}=\left(\matrix{U & 0 \cr 0 & V\cr}\right)
\left(\matrix{c_{\varphi_{\mu}} & 0 & s_{\varphi_{\mu}} & 0 \cr 0
& c_{\varphi_{\tau}} & 0 & s_{\varphi_{\tau}}\cr
-s_{\varphi_{\mu}}& 0 &c_{\varphi_{\mu}}& 0\cr 0
&-s_{\varphi_{\tau}}& 0 & c_{\varphi_{\tau}}\cr}\right),$$
$$U=\left(\matrix{c_{\theta_{\nu}} & e^{i\beta_{\nu}}s_{\theta_{\nu}} \cr
-e^{-\beta_{\nu}}s_{\theta_{\nu}}& c_{\theta_{\nu}}\cr}\right),
\qquad V=\left(\matrix{ c_{\theta_N} & e^{i\beta_N}s_{\theta_N}\cr
-e^{-i\beta_N}s_{\theta_N}& c_{\theta_N}\cr}\right),\eqno(8)$$
$\varphi_l$ is the mixing angle between light and heavy
neutrinos in the l-generation, $\theta_{\nu}$  ($\theta_N$) is the
mixing angle between light (heavy) neutrinos belonging to
different generations, $\beta_{\nu} (\beta_N)$ is the
$CP$-violating phase for light (heavy) neutrinos,
$s_{\varphi_{\mu}}= \sin\varphi_{\mu},\
c_{\varphi_{\mu}}=\cos\varphi_{\mu}$, and so on.

In the LRM the charged current Lagrangian is of the form
$${\cal{L}}^L_{CC}={g_L\over2\sqrt{2}}\overline{l}(x)\gamma_{\lambda}(1+\gamma_5)\nu_l(x)W_{L\lambda}(x)
+{g_R\over2\sqrt{2}}\overline{l}(x)\gamma_{\lambda}(1-\gamma_5)N_l(x)W_{R\lambda}(x),\eqno(9)$$
where
$$W_{L\lambda}=W_{1\lambda}c_{\xi}+W_{2\lambda}e^{-i\phi}s_{\xi},\qquad W_{R\lambda}=-W_{1\lambda}e^{i\phi}s_{\xi}+
W_{2\lambda}c_{\xi}$$ $\xi$ is the mixing angle in the sector of
the charged gauge bosons and $\phi$ is the phase to be responsible
for the $CP$ violation.

\begin{figure}[h]
\begin{center}
\includegraphics[]{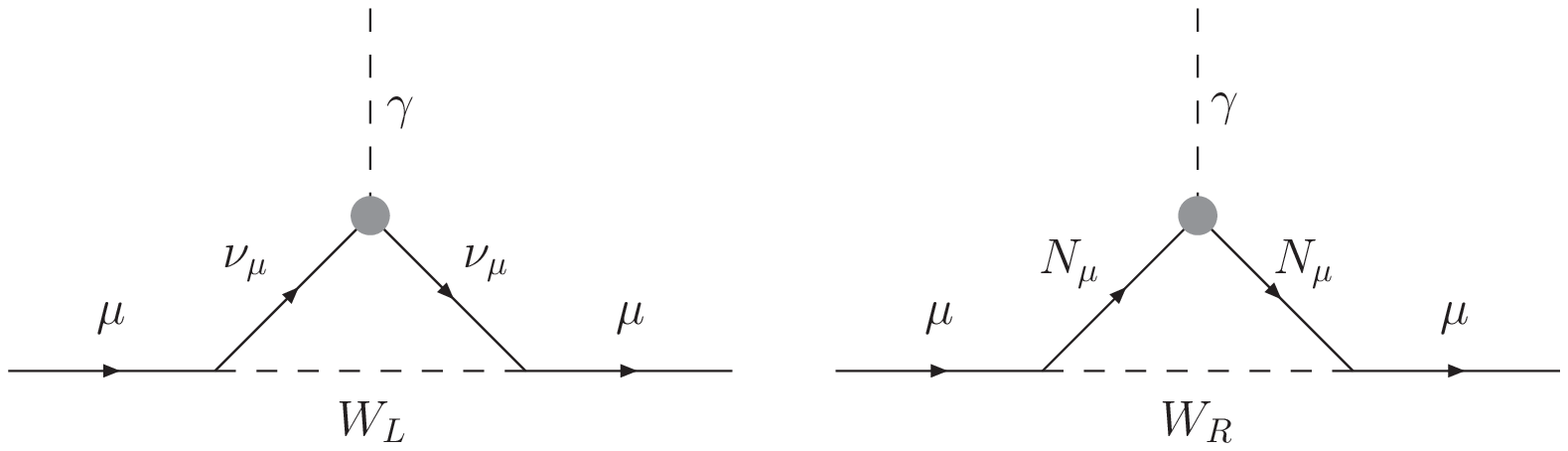}
\end{center}
\caption{One-loop diagrams contributing to the muon AMM due to the
light and heavy neutrinos.}
\end{figure}

In Fig. 1 we represent Feynman diagrams with light and heavy
neutrinos making a contribution into the muon AMM. The vertex
function of the third order $\Lambda_{\beta}(p,p^{\prime})$ that
corresponds to the diagrams with the $W^{(-)}_1$-boson exchange
has the form
$$\Lambda_{\beta}(p,p^{\prime})={eg^2_Lc^2_{\xi}\over(4\pi)^4m_e}\sum_{i,j}\mu^{\prime}_{ij}\int
\gamma_{\sigma}(1+\gamma_5)\left\{[i(\hat{p}^{\prime}-\hat{k})-m_i]\sigma_{\beta\tau}q_{\tau}
[i(\hat{p}-\hat{k})-m_j]{\cal{M}}^{\dagger}_{i\mu} {\cal{M}}_{\mu
i}\times\right.$$
$$\left.\times{\cal{M}}^{\dagger}_{j\mu}{\cal{M}}_{\mu j}+
C\lambda_{i\odot}m_i\sigma_{\beta\tau}q_{\tau}
C\lambda_{j\odot}m_j{\cal{M}}_{i\mu} {\cal{M}}_{\mu
i}{\cal{M}}^{\dagger}_{j\mu}{\cal{M}}^{\dagger}_{\mu
j}\right\}\gamma_{\nu}(1+\gamma_5)\times$$
$$\times{[\delta_{\sigma\nu}+k_{\sigma}k_{\nu}/m_{W_1}^2]d^4k\over
[k^2+m^2_{W_1}][(p^{\prime}-k)^2+m_i^2][(p-k)^2+m_j^2]},\eqno(10)$$
where $\mu^{\prime}_{ij}=\mu_{ij}/\mu_B$, $\mu_B$ is the Bohr
magneton, $\lambda_{j\odot}$ is the creation phase factor of
$\nu_j$-neutrino ($|\lambda_{j\odot}|^2=1$), $C$ is the charge
conjugation matrix, $i,j=\nu_2,\nu_3, N_2, N_3$. Besides, it is taken
into account that for the Majorana field the relations
$$<0|T(\nu_{\alpha}(x)\overline{\nu}_{\beta}(y))|0>=-iS^c_{\alpha\beta}(x-y),\qquad
<0|T(\nu_{\alpha}(x)\nu_{\beta}(y))|0>=-i\lambda_{\odot}mC_{\alpha\beta}\Delta^c(x-y)\eqno(11)$$
are valid. It is easy to verify that the second term in the
integrand vanishes. As a result, we arrive at the same expression
as in the case of the Dirac neutrino.

Of course, since the spontaneously broken gauge theories are
renormalizable, the vertex function
$\Lambda_{\beta}(p,p^{\prime})$ should be finite. In fact, the
total bare Lagrangian of the LRM includes no interactions of the
form given by Eq.(6). In the unitary gauge the propagator of the
$W_{1,2}^{(\pm)}$-bosons contains
$k_{\sigma}k_{\nu}/m_{W_{1,2}}^2$ term, resulting in a linearly
divergent piece. Although these divergent terms are canceled, the
finite parts depend on the routing of the momenta through the
graphs. Such an ambiguity could be resolved in several ways.

For example, one could employ the method presented in \cite{RJ},
where the $W_{1,2}^{(\pm)}$-boson lines carry no external momenta,
with the use of the $\xi$-limiting procedure. The second way is to
calculate the matrix elements in the $R_{\xi}$ gauge. Here we
choose more simple way, namely, we compute directly in the unitary
gauge, where the particle content is evident, and use the Dyson
procedure \cite{FD} that rests on the expansion in external
momenta of the subintegral expression with subsequent subtraction
of the divergent terms.

Making the vertex functions under investigation convergent and
singling out the part to be proportional to
$\gamma_{\lambda}\hat{q}-\hat{q}\gamma_{\lambda}$, we obtain an
expression for the contribution made by the neutrino DMM into the
muon AMM as follows:
$$\left(\delta a_{\mu}\right)_{nd}=\left(\delta a_{\mu}^{W_L}\right)_{nd}
+\left(\delta a_{\mu}^{W_R}\right)_{nd},\eqno(12)$$ where
$$\left(\delta a_{\mu}^{W_L}\right)_{nd}={G_Fm_{W_1}^2\over16\sqrt{2}
\pi^2m_e}\left(J_{W_1}^L+J_{W_2}^L\right),\qquad \left(\delta
a_{\mu}^{W_R}\right)_{nd}={G_Fm_{W_1}^2\over16\sqrt{2}
\pi^2m_e}\left(J_{W_1}^R+J_{W_2}^R\right),\eqno(13)$$
$$J^L_{W_1}=c^2_{\varphi_{\mu}}c_{\theta_{\nu}}s_{\theta_{\nu}}\left[c_{\xi}^2(m_{\nu_2}+m_{\nu_3})
\mu^{\prime}_{\nu_2\nu_3}m_{\mu}^2m_{W_1}^{-2}\int_0^1\left(x(1-3x)\ln\bigg|{l^{W_1}_{\nu_2}\over
L^{W_1}_{\nu_2}}\bigg|+ \right.\right.$$
$$\left.\left.+{(1-3x)l^{W_1}_{\nu_3}\over m_{\nu_2}^2-m_{\nu_3}^2}\ln
\bigg|{l^{W_1}_{\nu_2}\over
l^{W_1}_{\nu_3}}\bigg|+{(3x-1)L^{W_1}_{\nu_3}+(1-x)(m_{W_1}^2-
m_{\mu}^2x^2)\over
m_{\nu_2}^2-m_{\nu_3}^2}\ln\bigg|{L^{W_1}_{\nu_2}\over
L^{W_1}_{\nu_3}}\bigg| \right)dx+(m_{\nu_2}\leftrightarrow
m_{\nu_3})\right]+$$
$$+c_{\varphi_{\mu}}s_{\varphi_{\mu}}c_{\theta_{\nu}}s_{\theta_N}
\bigg[(m_{\nu_3}\rightarrow m_{N_3},
\mu^{\prime}_{\nu_2\nu_3}\rightarrow \mu^{\prime}_{\nu_2N_3},)
\bigg]+
c_{\varphi_{\mu}}s_{\varphi_{\mu}}c_{\theta_{\nu}}c_{\theta_N}\bigg[(m_{\nu_3}\rightarrow
m_{N_2},
\mu^{\prime}_{\nu_2\nu_3}\rightarrow\mu^{\prime}_{\nu_2N_2})\bigg]+$$
$$+s^2_{\varphi_{\mu}}c_{\theta_N}s_{\theta_N}\bigg[(m_{\nu_2}\rightarrow
m_{N_2}, m_{\nu_3}\rightarrow m_{N_3},
\mu^{\prime}_{\nu_2\nu_3}\rightarrow\mu^{\prime}_{N_2N_3}
)\bigg]+$$
$$+c_{\varphi_{\mu}}s_{\varphi_{\mu}}c_{\theta_N}
s_{\theta_{\nu}}\bigg[(m_{\nu_2}\rightarrow m_{N_2},
\mu^{\prime}_{\nu_2\nu_3}\rightarrow\mu^{\prime}_{N_2\nu_3})\bigg]+$$
$$+c_{\varphi_{\mu}}s_{\varphi_{\mu}}s_{\theta_{\nu}}s_{\theta_N}\bigg[(m_{\nu_2}\rightarrow m_{\nu_3},
m_{\nu_3}\rightarrow
m_{N_3},\mu^{\prime}_{\nu_2\nu_3}\rightarrow\mu^{\prime}_{\nu_3N_3})\bigg],\eqno(14)$$
$$J_{W_2}^L=J_{W_1}^L\left(\xi\rightarrow\xi+{\pi\over2},
m_{W_1}\rightarrow m_{W_2}\right),\qquad
J_{W_i}^R=J_{W_i}^L\left(\xi\rightarrow\xi+{\pi\over2},
\varphi_{\mu}\rightarrow\varphi_{\mu}+{\pi\over2}\right),$$
$$L^{W_k}_j=m_{\mu}^2x^2+(m_j^2-m_{\mu}^2-m_{W_k}^2)x+m_{W_k}^2,\qquad
l^{W_k}_j=(m_j^2-m_{W_k}^2)x+m_{W_k}^2,$$ $k=1,2,$ and we have set
$g_L=g_R$.

When analyzing the formula obtained, we must take into account
that the neutrino masses are not independent. In the two-flavor
approximation they are connected with the relation \cite{GO2}
$$\sum_j m_j=m_{\nu_2}+m_{\nu_3}+m_{N_2}+m_{N_3}=(f_{\mu\mu}+f_{\mu\tau})(v_L+v_R),\eqno(15)$$
where $f_{\mu\tau}$ and $f_{\mu\mu}$ are the triplet Yukawa
coupling constants, $v_{L,R}$ is the vacuum expectation values
(VEV) for the left- and right-handed triplets of the Higgs fields.

It is evident that the main contributions to $\delta a_{\mu}$ may
be due to the terms which are proportional to
$\mu^{\prime}_{N_2N_3}$ and $\mu^{\prime}_{N_a\nu_b}$ ($a,b=2,3$).
First, we consider the contribution coming from the terms
associated with $\mu^{\prime}_{N_2N_3}$ only. In our numerical
calculation we use the following parameter values:
$$\theta_{\nu}=\theta_N=50^0,\qquad
\xi=5\times10^{-2},\qquad m_{W_2}=800\ \mbox{GeV}.\eqno(16)$$

\begin{figure}[t]
\begin{center}
\includegraphics[scale=0.7]{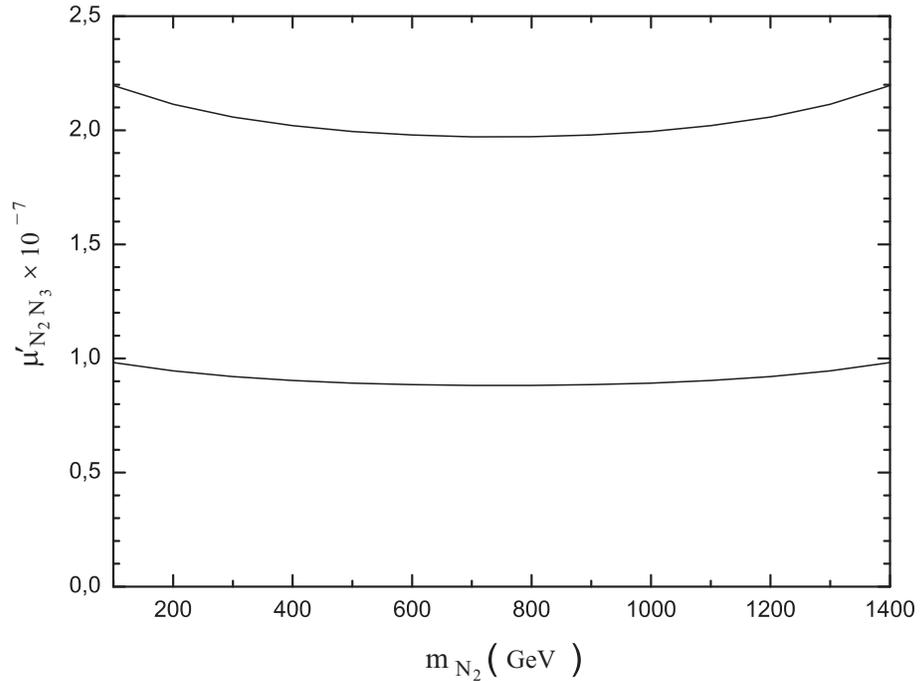}
\end{center}
\caption{The curves constraining the allowed region of the
$\mu^{\prime}_{N_2N_3}$ and $m_{N_2}$ values (the values of
$\mu^{\prime}_{N_2N_3}$ is taken in the Bohr magneton units).}
\end{figure}

For the case $\sum m_j=1500$ GeV and $\varphi_{\mu}=0$ in the
$\mu^{\prime}_{N_2N_3}$ vs $m_{N_2}$ parameter space, Fig. 2
presents two curves corresponding to the values of $\delta
a_{\mu}$ equal to $407\times10^{-11}$ (upper curve) and
$182\times10^{-11}$ (lower curve). The range of the parameters for
the heavy neutrino sector allowed by the BNL data lies between
these curves. It is seen that, in order to account for the
observed muon AMM, the values of $\mu_{N_2N_3}$ must be as large
as $10^{-7}\mu_B$.

Further we allow for the contributions made by $\mu_{N_a\nu_b}$.
In so doing, for the sake of simplicity, we assume that
$$\mu_{N_a\nu_b}=\mu_{N_2N_3}=\mu_{hn},$$
and take for $\varphi_{\mu}$ its upper bound $10^{-2}$ \cite{GO3}.
In that case we obtain that the $(g-2)_{\mu}$ anomaly finds its
explanation under fulfilment of the condition
$$\mu_{hn}\geq3\times10^{-8}\mu_B.$$

Let us examine, whether these values are at variance with the
literature data. Note that, up to now, there have been no
experimental limitations on the values of the heavy neutrino DMMs
(only the laboratory constraints for light neutrino DDMs have been
obtained). Therefore we should appeal to the values the theory
predicts.

Within the LRM, the transition DMMs have been computed in Refs.
\cite{RES}. In doing so the contributions of the diagrams with the
virtual charged gauge bosons were considered only. The diagrams
produced by the Lagrangians describing the interactions of the
singly charged Higgs bosons $h^{(-)}, \tilde{\delta}^{(-)}$ with
leptons and gauge bosons were disregarded. The authors of Refs.
\cite{RES} have neglected the mixing between the light and heavy
neutrinos too. The transition moments to be of interest here are
defined as follows:
$$\mu_{N_aN_b}\approx{-3eg_L^2(m_{N_a}+m_{N_b})\over2^7\pi^2}\left(
{s^2_{\xi} m_{\tau}^2\over m_{W_1}^4}+ {c^2_{\xi}m_{\tau}^2\over
m_{W_2}^4}\right)\mbox{Im}\left[V^*_{\tau a}V_{\tau
b}\right],\eqno(17)$$
$$\mu_{N_a\nu_b}\approx{eg_L^2\over8\pi^2}s_{\xi}c_{\xi}{m_{\tau}\over m_{W_1}^2}
\mbox{Im}\left[e^{-i\phi}V^*_{\tau a}U_{\tau b}\right].\eqno(18)$$

First of all, we note that in order to give positive contributions
to $\delta a_{\mu},$ the mixing matrices $U$, $V$, and
$CP$-violating phase $\phi$ must satisfy the conditions
$$\mbox{Im}\left[V^*_{\tau a}V_{\tau b}\right]<0,\qquad \mbox{Im}\left[e^{-i\phi}
V^*_{\tau a}U_{\tau b}\right]>0.\eqno(19)$$

In the case when the heavy neutrino mass is on the order of a few
TeV, from Eqs. (17) and (18) it follows that
$$\mu_{N_aN_b}\leq10^{-13}\mu_B, \qquad
\mu_{N_a\nu_b}\leq7\times10^{-11}\mu_B,\eqno(20)$$ where we have
made use of (16). So, for the case of the Majorana neutrino the
transition moment values required for the explanation of the
$(g-2)_{\mu}$ anomaly happen to be much larger than those predicted
by the theory.

As long as the values of $J_{W_i}^{L,R}$ as well as of the neutrino
DMMs are proportional to the neutrino mass, one might hope
to obtain the $(g-2)_{\mu}$ explanation at larger values of
$m_{N_{2,3}}$ (for example, at
$m_{N_2}+m_{N_3}>\mbox{few}\times10$ TeV). But, an increase in the
heavy neutrino mass is mainly caused by increasing size of
$v_R$. On the other hand, the quantity $v_R$ enters the definitions
both of the $W_2$ gauge boson mass and mixing angle $\xi$
$$m^2_{W_2}={g_L^2\over2}\left[(k_1^2+k_2^2+v_L^2+v_R^2)-\sqrt{(v_R^2-v_L^2)^2+k_1^2k_2^2}\right],$$
$$\tan2\xi={2k_1k_2\over v_R^2-v_L^2},$$
where $k_1$ and $k_2$ are VEV of neutral components of the
Higgs bidoublet. We recall that
$$v_L\ll\mbox{max}(k_1,k_2)\ll v_R.$$
Thus, with increase in a heavy neutrino mass, the $W_2$ gauge
boson mass is growing, whereas the mixing angle $\xi$ decreases. As a
result, the transition moment size for the heavy neutrino and also its
contribution to the muon AMM is diminishing.

In LRM neutrino could be of the Dirac nature as well. Then, the
theory predicts that the diagonal elements of the neutrino DMMs
are nonzero and could be much greater than the
nondiagonal elements. When the heavy neutrino masses lie on the
TeV scale, the heavy neutrino DMM is defined by the expression
\cite{RES}
$$\mu_{N_a}\approx{3eg_L^2m_{N_a}\over64\pi^2}\left[{s_{\xi}^2\over
m_{W_1}^2}+{c_{\xi}^2\over m_{W_2}^2}\right].\eqno(20)$$ Setting,
for example, $\xi=5\times10^{-2}$, $m_{W_2}=800$ GeV and
$m_{N_a}=1400$ GeV, we obtain
$$\mu_{N_a}\approx4\times10^{-9}\mu_B.\eqno(21)$$
So it may be hoped that big values of $\mu_{N_a}$ ensure the
$(g-2)_{\mu}$ anomaly explanation.

Taking into consideration the contribution to be made by the
diagrams with the heavy neutrino exchange, we arrive at
$$\delta a_{\mu}\approx{G_Fm_{W_1}^2c^2_{\varphi_{\mu}}m_{\mu}^2\over16\sqrt{2}
\pi^2m_e}\left\{c^2_{\theta_N} m_{N_2}
\mu^{\prime}_{N_2}\left[\int_0^1s^2_{\xi}\left({x(1-x)^2-x^3m_{N_2}^2m_{W_1}^{-2}
\over(xm_{\mu}^2-m_{W_1}^2)(x-1)+xm_{N_2}^2}+\right.\right.\right.
$$
$$\left.+m_{W_1}^{-2}x(1-3x)
\ln\left|{m_{W_1}^2(1-x)+m_{N_2}^2x\over(m_{W_1}^2-xm_{\mu}^2)(1-x)+m_{N_2}^2x}
\right|+m_{W_1}^{-2}x^2\right)dx+$$
$$\left.\left.+(\xi\rightarrow\xi+{\pi\over2}, m_{W_1}\rightarrow
m_{W_2})\right]
+\left(\theta_N\rightarrow\theta_N+{\pi\over2},\mu^{\prime}_{N_2}\rightarrow\mu^{\prime}_{N_3},
m_{N_2}\rightarrow m_{N_3}\right)\right\}.\eqno(22)$$ However, the
integrals in Eq. (22) prove to be too small and, despite
possible big values of $\mu_{N_a}$, the value of the muon AMM is inexplicable
by the presence of the neutrino DMM.

\section{Conclusions}

The $(g-2)_{\mu}$ anomaly within the LRM has been discussed. We
have considered the contributions coming from neutrinos possessing
the DMMs. As shown, in the Majorana case, in order to explain the
BNL result, the neutrino DMM must be as large as
$\times10^{-7}\mu_B$. The obtained value proves to be three orders
of magnitude higher than that calculated in Ref. \cite{RES}. Such
a situation is maintained as a neutrino has the Dirac nature.
Then, taking into account the results of Refs. \cite{{GO1},{JP},
{OGB}} we come to the conclusion: among the additional particles
of the LRM only the Higgs bosons could give explanation for the
$(g-2)_{\mu}$ anomaly.

At present, a new BNL muon experiment is under discussion. It is
scheduled to improve the results by at least a factor of 2. We
note that the up-to-date theoretical error is somewhat greater
than the experimental one. It is fully dominated by the
uncertainty in the hadronic low energy cross section data defining
the hadronic vacuum polarization and, in part, by the uncertainty
in the hadronic light-by-light scattering contribution. Therefore,
the theoretical results associated with calculations of the
hadronic corrections should be improved as well. Progress in this
area is awaited owing to the lattice QCD techniques. Comparison of
the theoretical and experimental results will give a more precise
$(g-2)_{\mu}$ value that, in turn, will enable finding of more
trustworthy constraints on the Higgs sector parameters for the
LRM.

\end{document}